\documentclass[12pt]{article}

\usepackage{epsfig}
\input epsf
\usepackage{subfigure}

\usepackage{amssymb}

\usepackage{fancyhdr}
\begin{document}

\title{Inflation on Moduli Space and Cosmic Perturbations}

\author{
Kenji Kadota$^1$ and Ewan D. Stewart$^2$ \\
$^1$ {\em Department of Physics, University of California, Berkeley, CA 94720, USA} \\
$^2$ {\em Department of Physics, KAIST, Daejeon 305-701, South Korea}
}

\maketitle   
\begin{abstract}
We show that a moduli space of the form predicted by string theory, lifted by supersymmetry breaking, gives rise to successful inflation for large regions of parameter space without any modification or fine tuning. This natural realization of inflation relies crucially on the complex nature of the moduli fields and the multiple points of enhanced symmetry, which are generic features of moduli space but not usually considered in inflationary model building.

Our scenario predicts cosmic perturbations with an almost exactly flat spectrum for a wide range of scales with running on smaller, possibly observable, scales. The running takes the form of either an increasingly steep drop off of the spectrum, or a rise to a bump in the spectrum before an increasingly steep drop off.
\end{abstract}

\thispagestyle{fancy}
\rhead{KAIST-TH2003/09\\UCB-PTH-03/27}

\section{Introduction}
\label{intro}

While inflation \cite{Gliner,guth1} solves many cosmological problems, the realization of such an accelerated expansion phase in the early Universe in particle theory still remains an open question.
Modular cosmology has also attracted a lot of attention \cite{example,towards} as string theory generically predicts ubiquitous moduli fields, though the moduli tend to cause cosmological problems \cite{bs} rather than solve them.
We introduced a successful modular cosmology scenario in Ref.~\cite{smc} where the problems in modular cosmology, in addition to the persistent fine-tuning problems in inflationary model building \cite{vpp}, are solved in a simple and self-consistent manner.
In this paper we show that the unresolved problem of the simplest single modulus model of Ref.~\cite{smc} is naturally resolved by a more careful analysis of the expected form of the modulus' potential (see Section~\ref{ntheta}), and give a more complete analysis of the cosmic perturbations produced in the model.

The outline of our paper is as follows.
In Section~\ref{modulispace} we review the expected form of a modulus' potential.
In Section~\ref{inflation} we describe our scenario of inflation on moduli space and present our detailed potential, including a toy model which we use for numerical calculations.
In Section~\ref{perts} we calculate the cosmic perturbations produced in our model.
In Section~\ref{obs} we summarize the observational predictions of our model.
In Section~\ref{conc} we discuss and conclude.

\section{Moduli space}
\label{modulispace}

In the following we briefly review the general properties of moduli space and the expected form of a modulus' potential after supersymmetry breaking.

\subsection{Moduli}

Before any supersymmetry breaking effects are taken into account, string theory has many flat directions in field space along which the potential vanishes.
The fields parameterizing such flat directions are called moduli and its space is referred to as moduli space.

These flat directions should not be confused with supersymmetric flat directions which are generically lifted by supersymmetric non-renormalizable terms and hence are only approximately flat in the neighborhood of an appropriately chosen origin even in the limit of unbroken supersymmetry.

\subsection{Points of enhanced symmetry}

The occurrence of symmetry enhancement at special points in moduli space is one of the most generic and robust phenomena in string theory \cite{stringy,cosmy}.
The points will generically have Planckian separations and the moduli some discrete symmetry about each point.
One also expects some fields (matter, gauge, or moduli) to become massless at a point of enhanced symmetry and become massive away from it, being Higgsed by the modulus that parameterizes the distance from the point of enhanced symmetry.
At typical points in moduli space, the moduli have Planckian values relative to the points of enhanced symmetry and so the Higgsed fields disappear from the low energy effective theory.

\subsection{Supersymmetry breaking}

Once supersymmetry breaking is taken into account, away from points of enhanced symmetry, the moduli acquire a potential of the form
\begin{equation}\label{moduluspot}
V(\Phi) = M_\mathrm{susy}^4 \, \mathcal{F}\left(\frac{\Phi}{M_\mathrm{Pl}}\right)
\end{equation}
where $M_\mathrm{susy}$ is the supersymmetry breaking scale \footnote{We consider a region of moduli space in which the scale of supersymmetry breaking does not vary greatly. This is a consequence of only considering the dynamics after the Brustein-Steinhardt problem \cite{bs} has been resolved by having the last local bit of eternal inflation at the local supersymmetry breaking scale \cite{smc}.}, $\cal{F}$ is a generic function, and $M_\mathrm{Pl} \equiv 1/\sqrt{8\pi G}$ which is hereafter set to 1.
The mass scale of this potential is $m \sim M_\mathrm{susy}^2/M_\mathrm{Pl}$.

The discrete symmetries of the moduli about the points of enhanced symmetry make the points of enhanced symmetry extrema of the potential.
In the neighborhood of any given point of enhanced symmetry, the interaction of the moduli with the fields that become light at the point of enhanced symmetry renormalizes the moduli's potential as a function of the distance from the point of enhanced symmetry.

\section{Inflation on moduli space}
\label{inflation}

\subsection{Overview of the scenario}
\label{over}
Our successful modular cosmology scenario is discussed in our previous paper \cite{smc} and consists of, roughly speaking, four stages
\begin{enumerate}
\item
An eternally inflating universe \cite{guth} consisting of an ensemble of all the eternally inflating extrema of the potential throughout the field space of string theory \cite{cosmo} - this makes all eternally inflating points in the field space of string theory equally likely, or more precisely one cannot say any one is more likely than any other.
\item
Locally, the last eternal inflation occurred at $V \lesssim M_\mathrm{susy}^4$ - this avoids the Brustein-Steinhardt problem \cite{bs}.
\item
Post-eternal modular inflation - this produces the observed scale-invariant perturbations.
\item
Thermal inflation \cite{thermal} - this solves cosmological moduli problem.
\end{enumerate}
In this paper we consider a simple single modulus model for the third stage of post-eternal modular inflation.

\subsection{The potential}
\label{generalpot}

We consider a modulus $\Phi$ with general properties as reviewed in Section~\ref{modulispace} and two neighboring points of enhanced symmetry in $\Phi$-space denoted by $\Phi_\smallfrown$ and $\Phi_\asymp$.

We assume that $\Phi$ has $\mathbb{Z}_2$ symmetry at $\Phi = \Phi_\asymp$ and symmetry greater than $\mathbb{Z}_2$ at $\Phi = \Phi_\smallfrown$.
Then, once supersymmetry is broken, the potential near $\Phi = \Phi_\smallfrown$ will have the form
\begin{equation}
V = V_\smallfrown - m^2_\smallfrown |\Phi-\Phi_\smallfrown|^2 + \ldots
\end{equation}
and that near $\Phi = \Phi_\asymp$ will have the form
\begin{equation}\label{sad}
V = V_\asymp + \frac{1}{2} \left[ \mu_\asymp^2 (\Phi-\Phi_\asymp)^2 + \mbox{c.c.} \right]
+ m^2_\asymp |\Phi-\Phi_\asymp|^2 + \ldots
\end{equation}
where $V_\smallfrown \sim V_\asymp \sim m_\smallfrown^2 \sim m_\asymp^2 \sim |\mu^2_\asymp| \sim M_\mathrm{susy}^4$.
We assume that $m^2_\smallfrown > 0$ so that $\Phi = \Phi_\smallfrown$ is a maximum and that $-|\mu^2_\asymp| < m^2_\asymp < |\mu^2_\asymp|$ so that $\Phi = \Phi_\asymp$ is a saddle.
This explains our notation, $\smallfrown$ for the maximum and $\asymp$ for the saddle.
Note that the discrete symmetry greater than $\mathbb{Z}_2$ at $\Phi = \Phi_\smallfrown$ ensures that we have an approximate global U(1) symmetry for small $|\Phi-\Phi_\smallfrown|$.

So far, we have ignored the interactions of $\Phi$ with the new fields that generically appear, \mbox{i.e.} become light, at the points of enhanced symmetry.
These interactions will renormalize the parameters in $\Phi$'s potential as a function of the distance from the point of enhanced symmetry.
We expect these interactions to respect an approximate global U(1) symmetry about their respective points of enhanced symmetry and so renormalize $m^2_\smallfrown$ and $m^2_\asymp$ but not $\mu^2_\asymp$.
Taking the effects of these interactions into account, the potential near $\Phi = \Phi_\smallfrown$ is
\begin{equation}
V = V_\smallfrown - \tilde{m}^2_\smallfrown |\Phi-\Phi_\smallfrown|^2 + \ldots
\end{equation}
and that near $\Phi = \Phi_\asymp$ is
\begin{equation}
V = V_\asymp + \frac{1}{2} \left[ \mu_\asymp^2 (\Phi-\Phi_\asymp)^2 + \mbox{c.c.} \right]
+ \tilde{m}^2_\asymp |\Phi-\Phi_\asymp|^2 + \ldots
\end{equation}
where $\tilde{m}^2_\smallfrown$ and $\tilde{m}^2_\asymp$ are functions of $\beta_\smallfrown \ln|\Phi-\Phi_\smallfrown|$ and $\beta_\asymp \ln|\Phi-\Phi_\asymp|$ respectively. 
We expect $\beta_\smallfrown \sim \beta_\asymp \sim 10^{-1}$ for our assumed moderately weak couplings and assume that $\tilde{m}^2_\smallfrown$ and $\tilde{m}^2_\asymp$ decrease as one approaches their respective points of enhanced symmetry.

\subsubsection{Potential near $\Phi = \Phi_\smallfrown$}

Defining the radial and angular components of the modulus about $\Phi = \Phi_\smallfrown$ as
\begin{equation}\label{ra}
\Phi-\Phi_\smallfrown = \frac{1}{\sqrt{2}\,} \phi e^{i\theta}
\end{equation}
the potential near $\Phi = \Phi_\smallfrown$ is
\begin{equation}
V = V_\smallfrown - \frac{1}{2} \tilde{m}^2_\smallfrown \phi^2 + \ldots
\end{equation}
where $\tilde{m}^2_\smallfrown$ is an increasing function of $\beta_\smallfrown\ln\phi$.
Following Ref.~\cite{fqc}, we define $\phi_0$ by
\begin{equation}
\left.\frac{dV}{d\phi}\right|_{\phi_0} = - \left(
\left.\tilde{m}^2_\smallfrown\right|_{\phi_0} + \frac{\beta_\smallfrown}{2}
\left.\frac{d\tilde{m}^2_\smallfrown}{d\beta_\smallfrown\ln\phi}\right|_{\phi_0}
\right) \phi_0 = 0
\end{equation}
We expect $\ln\phi_0 \sim - 1/\beta_\smallfrown$ and,
for consistency with the fact that the renormalization is cut off at a scale
$\sim M_\mathrm{susy}^2$, require $\phi_0 \gg M_\mathrm{susy}^2$.
Near $\phi = \phi_0$ we have
\begin{eqnarray}
V & = & V_\smallfrown - \frac{1}{2} \left[ \left.\tilde{m}^2_\smallfrown\right|_{\phi_0}
+ \left.\frac{d\tilde{m}^2_\smallfrown}{d\beta_\smallfrown\ln\phi}\right|_{\phi_0}
\beta_\smallfrown \ln\frac{\phi}{\phi_0}
+ \frac{1}{2} \left.\frac{d^2\tilde{m}^2_\smallfrown}{(d\beta_\smallfrown\ln\phi)^2}\right|_{\phi_0}
\beta_\smallfrown^2 \ln^2\frac{\phi}{\phi_0} + \ldots \right] \phi^2 \\
\label{maxloopgen}
& \simeq & V_\smallfrown - \frac{\beta_\smallfrown}{2}
\left.\frac{d\tilde{m}^2_\smallfrown}{d\beta_\smallfrown\ln\phi}\right|_{\phi_0}
\left( \ln\frac{\phi}{\phi_0} - \frac{1}{2} \right) \phi^2
\end{eqnarray}
\begin{equation}\label{Vpm}
\frac{dV}{d\phi} = - \beta_\smallfrown
\left.\frac{d\tilde{m}^2_\smallfrown}{d\beta_\smallfrown\ln\phi}\right|_{\phi_0}
\ln\frac{\phi}{\phi_0} \, \phi
\end{equation}
and
\begin{equation}\label{factorbeta}
\frac{d^2V}{d\phi^2} = - \beta_\smallfrown
\left.\frac{d\tilde{m}^2_\smallfrown}{d\beta_\smallfrown\ln\phi}\right|_{\phi_0}
\left( \ln\frac{\phi}{\phi_0} + 1 \right)
\end{equation}
We see that the loop corrections turn $\Phi = \Phi_\smallfrown$ into a local minimum and displace the maximum out to a ring at $\phi = \phi_0$.
The negative mass squared at this maximum is
\begin{equation}\label{m0}
m^2_0 = \beta_\smallfrown
\left.\frac{d\tilde{m}^2_\smallfrown}{d\beta_\smallfrown\ln\phi}\right|_{\phi_0}
\end{equation}
Note that this is suppressed by the factor $\beta_\smallfrown$ relative to our typical mass scales \cite{fqc}.

Eq.~(\ref{maxloopgen}) is very important as it will determine our predictions for the spectrum.
We see that it is accurate up to corrections of order $\beta_\smallfrown \ln(\phi/\phi_0)$.

\subsubsection{Potential near $\Phi = \Phi_\asymp$}
\label{pns}

Defining the real and imaginary components of the modulus about $\Phi = \Phi_\asymp$ as
\begin{equation}\label{ri}
\Phi-\Phi_\asymp = \frac{1}{\sqrt{2}\,} \left( \psi + i \chi \right)
\end{equation}
and without loss of generality taking $\mu_\asymp^2 > 0$,
the potential near $\Phi = \Phi_\asymp$ is
\begin{equation}
V = V_\asymp + \frac{1}{2} \mu_\asymp^2 \left( \psi^2 - \chi^2 \right)
+ \frac{1}{2} \tilde{m}^2_\asymp \left( \psi^2 + \chi^2 \right) + \ldots
\end{equation}
where $\tilde{m}^2_\asymp$ is an increasing function of $\beta_\asymp\ln(\psi^2+\chi^2)$.
We define $\psi_0$ by
\begin{equation}
\left.\frac{\partial V}{\partial\psi}\right|_{\psi=\psi_0, \chi=0} =
\left( \mu_\asymp^2 + \left.\tilde{m}^2_\asymp\right|_{\psi=\psi_0, \chi=0}
+ \beta_\asymp \left.\frac{d\tilde{m}^2_\asymp}{d\beta_\asymp\ln(\psi^2+\chi^2)}
\right|_{\psi=\psi_0, \chi=0} \right) \psi_0 = 0
\end{equation}
We expect $\ln\psi_0 \sim - 1/\beta_\asymp$ and,
for consistency with the fact that the renormalization is cut off at a scale
$\sim M_\mathrm{susy}^2$, require $\psi_0 \gg M_\mathrm{susy}^2$.
Near $\psi = \psi_0, \chi = 0$ we have
\begin{eqnarray}
V & = & V_\asymp + \frac{1}{2} \mu_\asymp^2 \left( \psi^2 - \chi^2 \right)
\nonumber \\ && \mbox{}
+ \frac{1}{2} \left[ \left.\tilde{m}^2_\asymp\right|_{\psi=\psi_0, \chi=0}
+ \left.\frac{d\tilde{m}^2_\asymp}{d\beta_\asymp\ln(\psi^2+\chi^2)}
\right|_{\psi=\psi_0, \chi=0} \beta_\asymp \ln\frac{\psi^2+\chi^2}{\psi_0^2}
+ \ldots \right] \left(\psi^2+\chi^2\right) \nonumber \\ \\
\label{sadloopgen}
& = & V_\asymp - \mu_\asymp^2 \chi^2
+ \frac{\beta_\asymp}{2} \left.\frac{d\tilde{m}^2_\asymp}{d\beta_\asymp\ln(\psi^2+\chi^2)}
\right|_{\psi=\psi_0, \chi=0} \left( \ln\frac{\psi^2+\chi^2}{\psi_0^2} - 1 \right)
\left(\psi^2+\chi^2\right)
\end{eqnarray}
\begin{equation}
\frac{\partial V}{\partial\psi} =
\beta_\asymp \left.\frac{d\tilde{m}^2_\asymp}{d\beta_\asymp\ln(\psi^2+\chi^2)}
\right|_{\psi=\psi_0, \chi=0} \ln\frac{\psi^2+\chi^2}{\psi_0^2} \, \psi
\end{equation}
and
\begin{equation}
\frac{\partial V}{\partial\chi} = - 2 \mu_\asymp^2 \chi
+ \beta_\asymp \left.\frac{d\tilde{m}^2_\asymp}{d\beta_\asymp\ln(\psi^2+\chi^2)}
\right|_{\psi=\psi_0, \chi=0} \ln\frac{\psi^2+\chi^2}{\psi_0^2} \, \chi
\end{equation}
We see that the loop corrections turn $\Phi = \Phi_\asymp$ into a local maximum with a saddle on either side at $\psi = \pm \psi_0$, $\chi=0$.
As $\chi$ rolls from a saddle towards the minimum, its mass squared runs from $-2\mu^2_\asymp$ to $-\mu^2_\asymp + m^2_\asymp$.
The exact form of the running between these two values will be model dependent.

\subsubsection{Toy potential}
\label{toy}

In this section we introduce a specific toy model which can be analyzed numerically.
It has zero $\Phi$-space curvature and bare potential
\begin{equation}\label{toymodulus}
V = V_\smallfrown - m^2_\smallfrown |\Phi-\Phi_\smallfrown|^2
+ \frac{1}{3} A m^2_\smallfrown \left[ (\Phi-\Phi_\smallfrown)^3 + \mbox{c.c.} \right]
+ \frac{1}{2} \nu (\nu+1) A^2 m^2_\smallfrown |\Phi-\Phi_\smallfrown|^4
\end{equation}
where our choice of parameterization is chosen for analytical convenience.
In this toy potential the symmetry at $\Phi = \Phi_\smallfrown$ is $\mathbb{Z}_3$ and that at $\Phi = \Phi_\asymp$ is a $\mathbb{Z}_2$ reflection symmetry.
We list some of its basic properties in the following.

For the vacuum energy to vanish at the minimum we require
\begin{equation}
A = \left(\frac{3\nu+1}{\nu^3}\right)^\frac{1}{2}
\left(\frac{m^2_\smallfrown}{6V_\smallfrown}\right)^\frac{1}{2}
\end{equation}
The minimum occurs at
\begin{equation}
\Phi_\smallsmile-\Phi_\smallfrown = \left(\frac{\nu}{3\nu+1}\right)^\frac{1}{2}
\left(\frac{6V_\smallfrown}{m^2_\smallfrown}\right)^\frac{1}{2} e^{i\pi/3}
\end{equation}
plus $\mathbb{Z}_3$ symmetric points, and the saddle is at 
\begin{equation}
\Phi_\asymp-\Phi_\smallfrown = \left(\frac{\nu}{\nu+1}\right)
\left(\frac{\nu}{3\nu+1}\right)^\frac{1}{2}
\left(\frac{6V_\smallfrown}{m^2_\smallfrown}\right)^\frac{1}{2}
\end{equation}
plus $\mathbb{Z}_3$ symmetric points.
The saddle parameters defined in Eq.~(\ref{sad}) are
\begin{equation}
V_\asymp = \left(\frac{1}{3\nu+1}\right) \left(\frac{2\nu+1}{\nu+1}\right)^3 V_\smallfrown
\end{equation}
\begin{equation}
m^2_\asymp = \left(\frac{\nu-1}{\nu+1}\right) m^2_\smallfrown
\ \ \ \mbox{and} \ \ \ 
\mu^2_\asymp = \left(\frac{\nu+2}{\nu+1}\right) m^2_\smallfrown
\end{equation}
As examples, we will take $\nu = 1, 2, 3$ as these values give very reasonable values of the physical parameters.
For example, $V_\asymp = \{0.84, 0.66, 0.54\} V_\smallfrown$ for $\nu = \{1, 2, 3\}$.

We take the loop corrections as
\begin{equation}
V_\mathrm{loop}^\smallfrown = - \beta_\smallfrown m^2_\smallfrown |\Phi-\Phi_\smallfrown|^2 \left( \ln\left|\frac{\Phi-\Phi_\smallfrown}{\Phi_\asymp-\Phi_\smallfrown}\right| - \frac{1}{2} \right)
\end{equation}
and
\begin{equation}
V_\mathrm{loop}^\asymp = \beta_\asymp m^2_\smallfrown |\Phi-\Phi_\asymp|^2
\left( \ln\left|\frac{\Phi-\Phi_\asymp}{\Phi_\smallfrown-\Phi_\asymp}\right|
- \frac{1}{2} \right)
\end{equation}
The $-\frac{1}{2}$'s are chosen to make the gradient of $V_\mathrm{loop}^\smallfrown$ zero at $\Phi = \Phi_\asymp$ and vice versa, which ensures that $V_\mathrm{loop}^\smallfrown$ does not interfere with the potential near $\Phi = \Phi_\asymp$ and vice versa.
This would be an automatic consequence of the symmetries in a real model.
For example, a $\mathbb{Z}_3$ symmetric $V_\mathrm{loop}^\asymp$ would be automatically flat at $\Phi = \Phi_\smallfrown$.
Our choice makes the model easier to handle numerically.

Defining the radial and angular components of the modulus about $\Phi = \Phi_\smallfrown$ as in Eq.~(\ref{ra}), the potential near $\Phi = \Phi_\smallfrown$ is
\begin{eqnarray}
V & = & V_\smallfrown
- \frac{1}{2} m^2_\smallfrown \left( 1 + \beta_\smallfrown \ln\frac{\phi}{\phi_\asymp}
- \frac{1}{2} \beta_\smallfrown \right) \phi^2 + \ldots \\
\label{maxlooptoy}
& = & V_\smallfrown - \frac{1}{2} m^2_0 \left( \ln\frac{\phi}{\phi_0} - \frac{1}{2} \right) \phi^2
\end{eqnarray}
where, for this toy model,
\begin{equation}
m^2_0 = \beta_\smallfrown m^2_\smallfrown
\end{equation}
and
\begin{equation}
\phi_0 = \phi_\asymp e^{-1/\beta_\smallfrown}
\end{equation}
Note that Eq.~(\ref{maxlooptoy}) has the same form as Eq.~(\ref{maxloopgen}).
Defining the real and imaginary components of the modulus about $\Phi = \Phi_\asymp$ as in Eq.~(\ref{ri}), the potential near $\Phi = \Phi_\asymp$ is
\begin{eqnarray}
\label{massrel}
V & = & V_\asymp + \frac{1}{2} \mu^2_\asymp \left( \psi^2 - \chi^2 \right)
+ \frac{1}{2} \left[ m^2_\asymp + \frac{1}{2} \beta_\asymp m^2_\smallfrown
\left( \ln\frac{\psi^2+\chi^2}{\psi_\smallfrown^2} - 1 \right) \right]
\left(\psi^2+\chi^2\right) + \ldots \\
& = & V_\asymp - \mu^2_\asymp \chi^2 + \frac{1}{4} \beta_\asymp m^2_\smallfrown
\left( \ln\frac{\psi^2 + \chi^2}{\psi_0^2} - 1 \right)
\left( \psi^2 + \chi^2 \right)
\end{eqnarray}
where
\begin{equation}
\psi_0 = \psi_\smallfrown \exp\left(-\frac{\mu^2_\asymp + m^2_\asymp}{\beta_\asymp m^2_\smallfrown}\right)
\end{equation}
Note that $\psi = \phi\cos\theta - \phi_\asymp$ and $\chi = \phi\sin\theta$ so that near $\Phi = \Phi_\asymp$ we have $\psi \simeq \phi - \phi_\asymp$ and $\chi \simeq \phi_\asymp\theta$.

\subsection{Inflationary dynamics}

{
\begin{figure}[ht]
\epsfxsize= 
3.9in
\begin{center}    
\leavevmode       
\epsffile{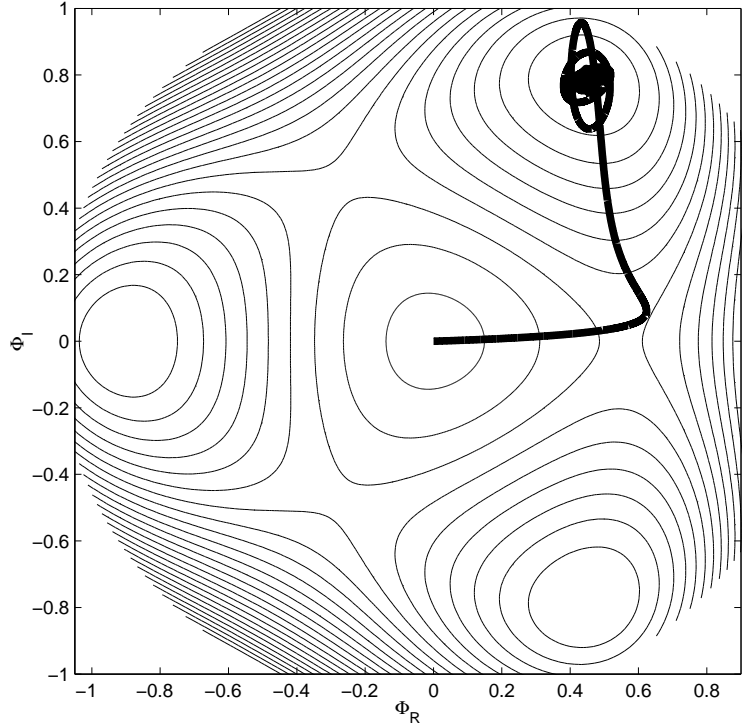}
\end{center}        
\caption{\label{potfig}
Trajectory of $\Phi = \left( \Phi_\mathrm{R} + i \Phi_\mathrm{I} \right) / \sqrt{2}$
for initial angle $\theta_\mathrm{i} = \pi/100$.}
\end{figure}
}

As outlined in Section~\ref{over}, our scenario starts with an eternally inflating universe consisting of an ensemble of eternally inflating extrema throughout the field space of string theory.
The initial conditions for our local part of the universe are taken to be
eternal inflation at the ring maximum $\phi = \phi_0$.
To ensure that the ring maximum satisfies the condition for eternal inflation \cite{smc},
$- V'' < 6 V$, we require
\begin{equation}\label{mcon1}
m^2_0 < 6 V_\smallfrown
\end{equation}
The flattening of the maximum by the factor $\beta_\smallfrown \ll 1$ shown in Eq.~(\ref{m0}) makes this a rather weak constraint.

The post-eternal modular inflation occurs as $\Phi$ rolls off the maximum, and observable scales should leave the horizon when $\Phi$ is still near $\phi = \phi_0$ so that $d\ln\phi/d\ln a \ll 1$.
In order not to get too many $e$-folds of inflation after this, we estimate \cite{fqc}
\begin{equation}\label{mcon2}
m^2_0 \gtrsim 0.1 V_\smallfrown
\end{equation}
Again, as we expect
$d\tilde{m}^2_\smallfrown / d\beta_\smallfrown\ln\phi \gtrsim V_\smallfrown$
and do not expect $\beta_\smallfrown$ to be very small, this is a rather weak constraint.

$\Phi$ may then roll down to the minimum in a straightforward manner, or, if the initial angle is just right, may do so via a more complicated trajectory passing through the neighborhood of a saddle on its way down.
As we show in Section~\ref{ntheta}, this seemingly improbable trajectory can be probable when considered a postiori due to the extra $e$-folds of expansion it induces, and is a crucial part of our model.

The contour plot of potential with some background field trajectory is shown in Fig.~\ref{potfig}.

\section{Perturbations}
\label{perts}

\subsection{$\delta N$ formalism}
\label{formalism}

The $\delta N$ formalism is a general and intuitive method to calculate the final comoving curvature perturbation when there exist multiple light degrees of freedom in the early universe \cite{deltaN}.
It expresses the final curvature perturbation in terms of the perturbation in the number of $e$-folds of expansion from an initial spatially flat hypersurface to the final comoving, or equivalently constant energy density, hypersurface,
\begin{equation}
\mathcal{R}(t_\mathrm{f}) = \delta N
\end{equation}
When the multiple light degrees of freedom on the initial hypersurface are scalar fields $\phi_i$, their contributions to the final curvature perturbation can be expressed as
\begin{equation}
\mathcal{R}(t_\mathrm{f}) =
\sum_i \frac{\partial N}{\partial \phi_i} \, \delta\phi_i(t_\mathrm{i})
+ \sum_i \frac{\partial N}{\partial \dot\phi_i} \, \delta\dot\phi_i(t_\mathrm{i})
\end{equation}
where $\delta\phi_i$ and $\delta\dot\phi_i$ are evaluated on the initial flat hypersurface at super-horizon scales after any $k$ dependent evolution has become negligible.
If the growing mode dominates, $N(\phi_i,\dot\phi_i)$ can be re-expressed as $N(\phi_i)$, and the above can be simplified to
\begin{equation}
\mathcal{R}(t_\mathrm{f}) =
\sum_i \frac{\partial N}{\partial \phi_i} \, \delta\phi_i(t_\mathrm{i})
\end{equation}

The power spectrum is defined by
\begin{eqnarray}
\frac{2\pi^2}{k^3} \, \mathcal{P}_\mathcal{R}(k) \,
\delta^3(\mathbf{k}-\mathbf{l})
& \equiv & \left\langle \mathcal{R}(\mathbf{k},t_\mathrm{f}) \,
{\mathcal{R}(\mathbf{l},t_\mathrm{f})}^\dagger \right\rangle \\
& = & \sum_{i,j}
\frac{\partial N}{\partial \phi_i} \frac{\partial N}{\partial \phi_j}
\left\langle \delta\phi_i(\mathbf{k},t_\mathrm{i}) \,
{\delta\phi_j(\mathbf{l},t_\mathrm{i})}^\dagger \right\rangle
\end{eqnarray}
Therefore
\begin{equation}
\mathcal{P}_\mathcal{R}(k) = \sum_{i,j}
\frac{\partial N}{\partial \phi_i} \frac{\partial N}{\partial \phi_j} \,
\mathcal{P}_{\delta\phi_i \delta\phi_j^\dagger}(k,t_\mathrm{i})
\end{equation}
where
\begin{equation}
\frac{2\pi^2}{k^3} \,
\mathcal{P}_{\delta\phi_i \delta\phi_j^\dagger}(k,t_\mathrm{i}) \,
\delta^3(\mathbf{k}-\mathbf{l})
\equiv \left\langle \delta\phi_i(\mathbf{k},t_\mathrm{i}) \,
{\delta\phi_j(\mathbf{l},t_\mathrm{i})}^\dagger \right\rangle
\end{equation}
If the $\delta\phi_i$ are independent at $t_\mathrm{i}$
then this simplifies to
\begin{equation}
\mathcal{P}_\mathcal{R}(k) =
\sum_i \left( \frac{\partial N}{\partial \phi_i} \right)^2
\mathcal{P}_{\delta\phi_i}(k,t_\mathrm{i})
\end{equation}
where $\mathcal{P}_{\delta\phi_i} \equiv \mathcal{P}_{\delta\phi_i \delta\phi_i^\dagger}$.

In our case we have
\begin{equation}\label{formula}
\mathcal{P}_\mathcal{R} =
\left( \frac{\partial N}{\partial \phi} \right)^2 \mathcal{P}_{\delta\phi}
+ \left( \frac{\partial N}{\partial \theta} \right)^2 \mathcal{P}_{\delta\theta}
\end{equation}

\subsection{$\phi$}

Observable scales leave the horizon while $\phi$ is rolling off the ring maximum at $\phi = \phi_0$.
The approximate $U(1)$ symmetry will be accurate for $\phi \ll 1$ and hence $\theta$ will be constant.
For $\beta_\smallfrown \ln(\phi/\phi_0) \ll 1$, Eq.~(\ref{Vpm}) gives the equation of motion
\begin{equation}\label{phieom}
\ddot\phi + 3H_0\dot\phi - m^2_0 \ln\frac{\phi}{\phi_0} \, \phi = 0
\end{equation}
where $H_0 \equiv \sqrt{V_\smallfrown/3}$.

We can solve this equation analytically in two cases.
For $\ln(\phi/\phi_0) \ll 1$ we have
\begin{equation}
\ln\frac{\phi}{\phi_0} \simeq \frac{\phi-\phi_0}{\phi}
\end{equation}
and so
\begin{equation}
\phi - \phi_0 \propto e^{\alpha H_0 t}
\end{equation}
where
\begin{equation}\label{alpha}
\alpha = \frac{3}{2} \left(
\sqrt{ 1 + \frac{4 m^2_0}{3 V_\smallfrown} }\, - 1 \right)
\end{equation}
Alternatively, for $m^2_0 \ll V_\smallfrown$ and $\ln(\phi/\phi_0) \lesssim 1$ we can use the slow-roll approximation to give
\begin{equation}
\ln\frac{\phi}{\phi_0} \propto \exp\left( \frac{m^2_0}{V_\smallfrown} H_0 t \right)
\end{equation}
We can combine these as
\begin{equation}\label{phiapprox}
\ln\frac{\phi}{\phi_0} \propto e^{\alpha H_0 t}
\end{equation}
which is accurate in both limits,
\mbox{i.e.} for
\begin{equation}\label{limit}
\alpha \ln\frac{\phi}{\phi_0} \ll 1
\end{equation}
Eq.~(\ref{phiapprox}) gives
\begin{equation}
\frac{d\ln\phi}{d\ln a} = \alpha \ln\frac{\phi}{\phi_0}
\end{equation}
so the condition Eq.~(\ref{limit}) is equivalent to
\begin{equation}\label{phislow}
\frac{d\ln\phi}{d\ln a} \ll 1
\end{equation}

\subsection{$\delta\phi$}
\label{gcdphi}

To calculate the contribution of the fluctuations $\delta\phi$ to the final curvature perturbation, we simply note that the angular fluctuations decouple and hence we can use the standard single field results.
For $\ln(\phi/\phi_0) \ll 1$ the results of Ref.~\cite{more} give
\begin{equation}
\label{note}
\left( \frac{\partial N}{\partial \phi} \right)^2 \mathcal{P}_{\delta\phi}
= \left[\frac{H_0}{2\pi\alpha(\phi_\star-\phi_0)}\right]^2
\left[ 2^\alpha \frac{\Gamma(\alpha+\frac{3}{2})}{\Gamma(\frac{3}{2})} \right]^2
\left(\frac{k}{a_\star H_0}\right)^{-2\alpha}
\end{equation}
where $\star$ is an arbitrary evaluation point satisfying $\ln(\phi_\star/\phi_0) \ll 1$.

\subsection{$\delta\theta$}
\label{gcd}

To calculate $\delta\theta$,
on super-horizon scales but while the angular potential is still flat,
we need to solve the perturbed equation of motion for $\delta\theta(k,t)$
\begin{equation}\label{deltathetaeom}
\frac{d}{dt} \left( a^3 \phi^2 \frac{d\,\delta\theta}{dt} \right)
+ a k^2 \phi^2 \delta\theta = 0
\end{equation}
Defining $\varphi \equiv a \phi \, \delta\theta$,
the conformal time $d\eta \equiv dt/a$ and $x \equiv -k \eta$,
we can write this as
\begin{equation}
\frac{d^2\varphi}{dx^2}
+ \left[ 1 - \frac{1}{a\phi} \frac{d^2(a\phi)}{dx^2} \right] \varphi = 0
\end{equation}
Following Refs.~\cite{smc,gsr} and introducing
\begin{equation}
f \equiv \frac{2\pi x a \phi}{k}
\end{equation}
for
\begin{equation}\label{fcon}
\frac{d\ln f}{d\ln x} \ll 1
\end{equation}
we have \cite{gsr}
\begin{equation}
\ln \mathcal{P}_{\delta\theta} = \ln\frac{1}{f_\star^2}
- 2 \int_0^\infty \frac{dx}{x} \left[\omega(x)-\theta(x_\star-x)\right]
\frac{d\ln f}{d\ln x}
\end{equation}
where the window function $\omega(x)$ is
\begin{equation}
\omega(x) = \frac{\sin(2x)}{x} - \cos(2x)
= 1 + \frac{2}{3} x^2 + \mathcal{O}\left(x^4\right)
\end{equation}
$\theta(x) = 0$ for $x<0$ and $\theta(x) = 1$ for $x>0$,
and $\star$ is an arbitrary evaluation point.
If $f \rightarrow f_\infty$ as $x \rightarrow \infty$, we can write this as
\begin{equation}\label{Pdth}
\ln\mathcal{P}_{\delta\theta} = \ln\frac{1}{f_\infty^2}
+ \int_0^\infty \frac{dx}{x} \left[ - x \, \omega'(x) \right] \ln\frac{f_\infty^2}{f^2}
\end{equation}
Note that
\begin{equation}
\int_0^\infty \frac{dx}{x} \left[ - x \, \omega'(x) \right] = 1 + \cos(2\infty)
\end{equation}
and
\begin{equation}\label{neg}
\lim_{x \rightarrow 0} \left[ - x \, \omega'(x) \right] =
- \frac{4}{3} x^2 + \mathcal{O}\left(x^4\right)
\end{equation}
thus the super-horizon part gives a negative contribution.

For $H = H_0$, we have
\begin{equation}
x = \frac{k}{a H_0}
\ , \ \ 
f = \frac{2\pi\phi}{H_0}
\ , \ \ 
f_\infty = \frac{2\pi\phi_0}{H_0}
\end{equation}
and the condition Eq.~(\ref{fcon}) becomes equivalent to the condition Eq.~(\ref{phislow}) and hence the condition Eq.~(\ref{limit}).
Substituting into Eq.~(\ref{Pdth}) gives
\begin{equation}\label{paeig}
\ln \mathcal{P}_{\delta\theta} = 2 \ln\frac{H_0}{2\pi\phi_0}
- 2 \int_0^\infty \frac{dx}{x} \left[ - x \, \omega'(x) \right] \ln\frac{\phi}{\phi_0}
\end{equation}
For $\alpha \ln(\phi/\phi_0) \ll 1$, Eq.~(\ref{phiapprox}) gives
\begin{equation}\label{lnphi}
\ln\frac{\phi}{\phi_0} =
\ln\frac{\phi_\star}{\phi_0} \left(\frac{x_\star}{x}\right)^\alpha
\end{equation}
and if that is the only relevant part of the integral then
\begin{eqnarray}\label{paei}
\ln \mathcal{P}_{\delta\theta} & = & 2 \ln\frac{H_0}{2\pi\phi_0}
- 2 \int_0^\infty \frac{\left[ - x \, \omega'(x) \right] dx}{x^{1+\alpha}}
\ln\frac{\phi_\star}{\phi_0} \, x_\star^\alpha \\
\label{pae}
& = & 2 \ln\frac{H_0}{2\pi\phi_0} - 2 \left[
2^\alpha \cos\left(\frac{\pi\alpha}{2}\right) \frac{\Gamma(2-\alpha)}{1+\alpha} \right]
\ln\frac{\phi_\star}{\phi_0} \left(\frac{k}{a_\star H_0}\right)^\alpha
\end{eqnarray}
For $\alpha < 1$, Eq.~(\ref{pae}) fits well with our intuition from the slow-roll limit ($\alpha \ll 1$); $\phi$ is increasing so $\delta\theta \sim H_0 / \phi$ is decreasing.
For $\alpha > 1$, however, $\ln(\phi/\phi_0)$ is growing sufficiently fast that the negative super-horizon part of the window function shown in Eq.~(\ref{neg}) dominates the integral and we get a rising spectrum.
This is the reason for the sign change in Eq.~(\ref{pae}) at $\alpha = 1$.
For $\alpha \geq 2$, the growth of $\ln(\phi/\phi_0)$ is so fast that the integral in Eq.~(\ref{paei}) diverges indicating that effects well outside the horizon dominate \cite{misao}\footnote{E.D.S. thanks Misao Sasaki for explaining these ideas to him.}.
This is why Eq.~(\ref{pae}) diverges at $\alpha = 2$.

When $\alpha\ln(\phi/\phi_0)$ becomes of order one, our approximation breaks down, but also
the rapid growth of $\ln(\phi/\phi_0)$ cuts off.
For $\alpha > 1$, this means that the spectrum will turn down leaving us with a bump in the spectrum at values of $k$ corresponding to modes which left the horizon when $\alpha\ln(\phi/\phi_0) \sim 1$.
For $\alpha \geq 2$, it means that the integral in Eq.~(\ref{paeig}) does converge but is dominated by the point where the rapid growth of $\ln(\phi/\phi_0)$ is cut off, \mbox{i.e.} $\alpha\ln(\phi/\phi_0) \sim 1$.
We can crudely model this analytically by using Eq.~(\ref{lnphi}) but cutting off the integral at $x = x_\mathrm{c}$ corresponding to $\alpha\ln(\phi/\phi_0) \sim 1$.
This also improves the accuracy of the formula for $\alpha < 2$.
Our improved formula is
\begin{eqnarray}
\ln \mathcal{P}_{\delta\theta} & \sim & 2 \ln\frac{H_0}{2\pi\phi_0}
- 2 \int_{x_\mathrm{c}}^\infty \frac{\left[ - x \, \omega'(x) \right] dx}{x^{1+\alpha}}
\ln\frac{\phi_\star}{\phi_0} \, x_\star^\alpha \\
\label{apxc}
& \simeq & 2 \ln\frac{H_0}{2\pi\phi_0} - 2 \left[
2^\alpha \cos\left(\frac{\pi\alpha}{2}\right) \frac{\Gamma(2-\alpha)}{1+\alpha}
+ \frac{4 x_\mathrm{c}^{2-\alpha}}{3(2-\alpha)} \right]
\ln\frac{\phi_\star}{\phi_0} \left(\frac{k}{a_\star H_0}\right)^\alpha
\end{eqnarray}
where we have assumed that $x_\mathrm{c} \ll 1$ and used Eq.~(\ref{neg}).
For $\alpha = 2$ we have
\begin{equation}
2^\alpha \cos\left(\frac{\pi\alpha}{2}\right) \frac{\Gamma(2-\alpha)}{1+\alpha}
+ \frac{4 x_\mathrm{c}^{2-\alpha}}{3(2-\alpha)}
= \frac{4}{3} \left[ \ln(2x_\mathrm{c}) + \gamma - \frac{1}{3} \right]
\end{equation}
where $\gamma \simeq 0.5772$ is the Euler-Mascheroni constant.
Stretching Eq.~(\ref{lnphi}) beyond its range of validity, we have
\begin{equation}\label{axc}
x_\mathrm{c} \sim
\left( \alpha \ln\frac{\phi_\star}{\phi_0} \right)^\frac{1}{\alpha} x_\star
\end{equation}
Note that due to the cutoff the rise in the spectrum saturates at proportional to $k^2$ for $\alpha > 2$.

\begin{figure}[p]
\begin{center}
\epsfxsize = 0.7\textwidth
\epsffile{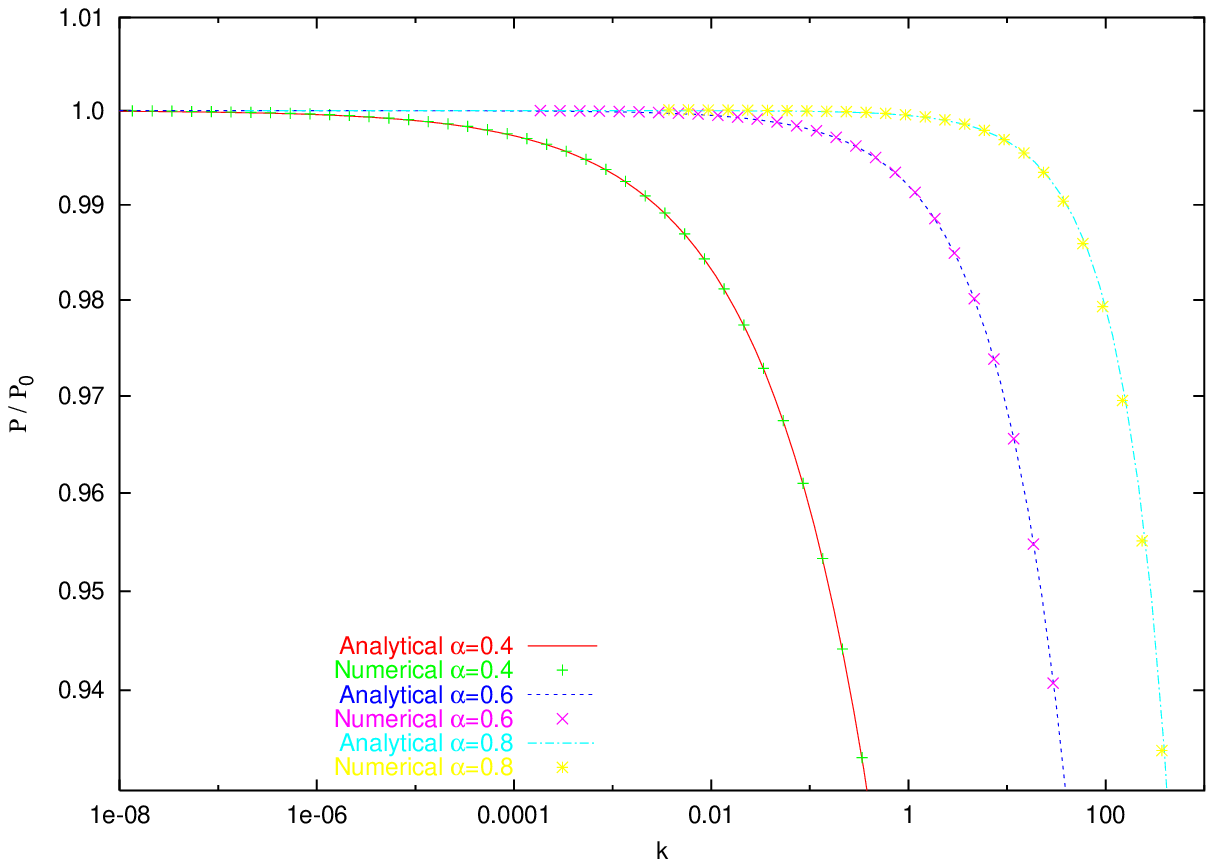}
\epsfxsize = 0.7\textwidth
\epsffile{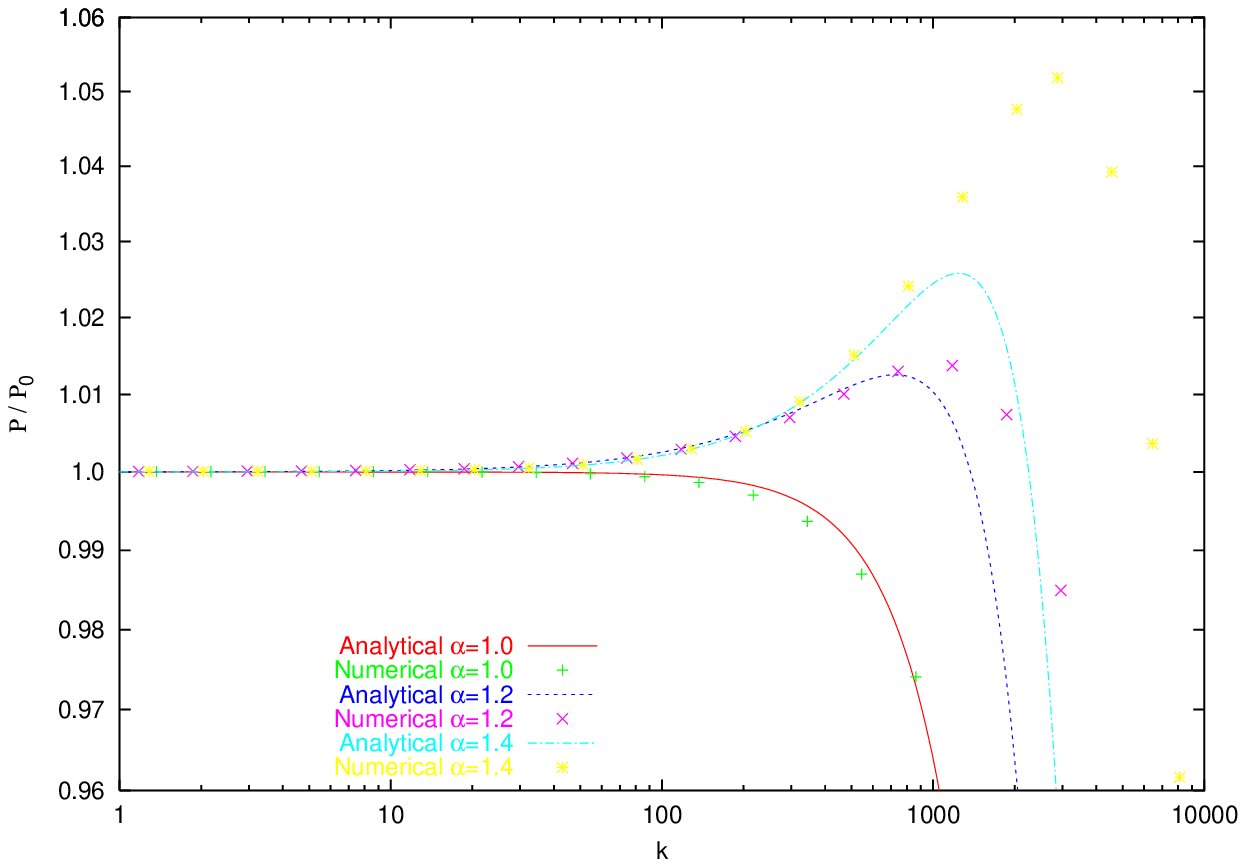}
\end{center}
\caption{\label{pnafig}
Log-log plots of $\left(2\pi\phi_0/H_0\right)^2 \mathcal{P}_{\delta\theta}$ as a function of $k$ for (top) $\alpha = 0.4, 0.6, 0.8$ and (bottom) $\alpha = 1.0, 1.2, 1.4$, from left to right respectively.
The points are numerical solutions of Eqs.~(\ref{phieom}) and~(\ref{deltathetaeom})
and the continuous lines are plots of the analytic formula of Eqs.~(\ref{apxc}) and~(\ref{axc}).
We fix $a$ and hence $k$ for the numerical plots by taking $aH = 10^4$ at $\alpha\ln(\phi/\phi_0) = 1$ so that the mode $k=10^4$ left the horizon at $\alpha\ln(\phi/\phi_0) = 1$.
We fix $a$ for the analytical plots by matching to the value of the corresponding numerical plot at $\alpha\ln(\phi/\phi_0) = 10^{-3}$.
}
\end{figure}

We have compared our analytical formula with numerical solutions of Eqs.~(\ref{phieom}) and~(\ref{deltathetaeom}).
Some of the results are shown in Fig.~\ref{pnafig}.
For $\alpha \leq 0.8$, the analytic formula works very well.
For $1.2 \leq \alpha \leq 1.8$, the analytic formula accurately reproduces the asymptotic behavior at small $k$ but poorly determines the shape of the bump.
This is to be expected as the bump occurs at $\alpha\ln(\phi/\phi_0) \sim 1$ where the analytic formula is expected to break down.
For $\alpha \geq 2$, the analytic formula reproduces the asymptotic scaling at small $k$ but cannot determine the coefficient accurately due to our crude use of the cutoff $x_\mathrm{c}$, and the shape of the bump is again poorly reproduced as we just discussed.
A full range of numerical solutions is given in Fig.~\ref{pfig} in Section~\ref{ps},
and a more accurate analytic treatment of the regime where super-horizon effects dominate will be given elsewhere.

\subsection{$\partial N/\partial \theta$}
\label{ntheta}

$\partial N/\partial \theta$ is the last ingredient we need in order to determine the final curvature perturbation spectrum using Eq.~(\ref{formula}).
$\partial N/\partial \theta$ will be independent of $k$ and $t_\mathrm{i}$, and so just provide the normalization of the contribution of $\delta\theta$ to the final curvature perturbation, assuming that we take $t_\mathrm{i}$ sufficiently early that the approximate U(1) symmetry about $\Phi = \Phi_\smallfrown$ is still accurate.

In Ref.~\cite{smc} we showed that $\partial N/\partial \theta$ has a strong dependence on the initial angle $\theta$.
A priori, the U(1) symmetry will make all angles equally likely and for most angles $\partial N/\partial \theta$ is of order one.
However, initial angles that roll down towards the saddle give extra inflation as the modulus rolls off the saddle and so produce a greater final volume.
These angles may then be preferred a postiori. Furthermore, as the amount of extra inflation is very sensitive to the initial angle,
we can expect $\partial N/\partial \theta$ to be very large for these angles.
Thus to determine expected values of $\partial N/\partial \theta$ we need to calculate the a postiori probability distribution of the initial angle $\theta$.

The a postiori probability density of $\theta$ is
\begin{equation}
P(\theta) \propto e^{3N(\theta)}
\end{equation}
However, taking $\theta = 0$ to be the initial angle that rolls down to the saddle, we expect to consider exponentially small values of $\theta$.
Therefore it will be more convenient to consider the probability density of $\ln\theta$
\begin{equation}
P(\ln\theta) \propto \theta e^{3N(\theta)}
\end{equation}
and hereafter $P$ will represent $P(\ln\theta)$.
Now
\begin{equation}
\frac{d P}{d\ln\theta} \propto
\left( 3 \frac{d N}{d \ln\theta} + 1 \right) \theta e^{3N}
\end{equation}
which has a non-trivial extremum at $\theta = \theta_\mathrm{c}$ where
\begin{equation}\label{Pext}
\left.\frac{d N}{d\ln\theta}\right|_{\theta_\mathrm{c}} = - \frac{1}{3}
\end{equation}
and
\begin{equation}
\frac{d^2 P}{(d\ln\theta)^2} \propto \left[ 3 \frac{d^2 N}{(d\ln\theta)^2}
+ \left( 3 \frac{d N}{d \ln\theta} + 1 \right)^2 \right] \theta e^{3N}
\end{equation}
If
\begin{equation}\label{Pmax}
\left.\frac{d^2N}{(d\ln\theta)^2}\right|_{\theta_\mathrm{c}} < 0
\end{equation}
the extremum is a maximum, and we can expect
\begin{equation}
\ln \theta \sim \ln \theta_\mathrm{c} \pm \sigma
\end{equation}
where
\begin{equation}\label{sigma}
\frac{1}{\sigma^2}
\equiv - \left.\frac{1}{P} \frac{d^2 P}{(d\ln\theta)^2}\right|_{\theta_\mathrm{c}}
= - 3 \left.\frac{d^2 N}{(d\ln\theta)^2}\right|_{\theta_\mathrm{c}}
\end{equation}
Thus, using Eq.~(\ref{Pext}), $\partial N/\partial \theta$ can be estimated as
\begin{equation}\label{dndth}
\frac{\partial N}{\partial\theta} \sim - \frac{e^{\pm\sigma}}{3\theta_\mathrm{c}}
\end{equation}

To determine $P(\ln\theta)$, and in particular $\theta_\mathrm{c}$ and $\sigma$, we need to solve the dynamics around the saddle.
Section~\ref{pns} gives the potential near the saddle as
\begin{equation}
V = V_\asymp + \frac{1}{2} \mu_\asymp^2 \left( \psi^2 - \chi^2 \right)
+ \frac{1}{2} \tilde{m}^2_\asymp \left( \psi^2 + \chi^2 \right) + \ldots
\end{equation}
where $\tilde{m}^2_\asymp$ runs from $-\mu_\asymp^2$ at the saddle to $m^2_\asymp$ towards the minimum.
Treating $V_{\chi\chi} = - \mu_\asymp^2 + \tilde{m}^2_\asymp$ as slowly varying, we get
\begin{equation}
\chi \propto e^{\int \alpha_\chi H dt}
\end{equation}
where
\begin{equation}\label{achi}
\alpha_\chi = \frac{3}{2}
\left( \sqrt{ 1 + \frac{4(\mu_\asymp^2-\tilde{m}^2_\asymp)}{3V_\asymp} }\, - 1 \right)
\end{equation}
and therefore
\begin{equation}
\frac{d\ln\chi}{d\ln a} = \alpha_\chi
\end{equation}
Similarly, averaging over any oscillations,
\begin{equation}
\psi \propto e^{- \int \alpha_\psi H dt}
\end{equation}
and
\begin{equation}
- \frac{d\ln\psi}{d\ln a} = \alpha_\psi
\end{equation}
where, since the imaginary part represents oscillations which we average out, 
\begin{equation}\label{apsi}
\alpha_\psi = \frac{3}{2} \, \mathrm{Re}
\left( 1 - \sqrt{ 1 - \frac{4(\mu_\asymp^2+\tilde{m}^2_\asymp)}{3V_\asymp} } \right)
\end{equation}

Letting subscript $\mathrm{e}$ denote the time when $\chi = \psi$, we have
\begin{eqnarray}\label{chii}
\chi_\mathrm{i}
& = & \psi_\mathrm{i} \exp \left( - \int_{\ln\psi_\mathrm{e}}^{\ln\psi_\mathrm{i}}
\frac{\alpha_\psi+\alpha_\chi}{\alpha_\psi} d\ln\psi \right) \\
\label{chii2}
& \simeq & \psi_\mathrm{i}
\exp \left( - \frac{1}{\beta_\asymp} \int_{\tilde{m}^2_{\asymp\mathrm{e}}}^{m^2_\asymp}
\frac{\alpha_\psi+\alpha_\chi}{2\alpha_\psi}
\left[\frac{d\tilde{m}^2_\asymp}{d\beta_\asymp\ln(\psi^2+\chi^2)}\right]^{-1}
d\tilde{m}^2_\asymp \right)
\end{eqnarray}
The number of $e$-folds from the initial time $t_\mathrm{i}$ until a final time $t_\mathrm{f}$, corresponding to some fixed final value $\chi = \chi_\mathrm{f}$, is
\begin{equation}
N = \int_{t_\mathrm{i}}^{t_\mathrm{f}} H dt
= \int_{\ln\chi_\mathrm{i}}^{\ln\chi_\mathrm{f}} \frac{d\ln\chi}{\alpha_\chi}
= \int_0^{\ln\frac{\chi_\mathrm{e}}{\chi_\mathrm{i}}}
\frac{1}{\alpha_\chi} \, d\ln\frac{\chi}{\chi_\mathrm{i}}
+ \int_{\ln\frac{\chi_\mathrm{e}}{\chi_\mathrm{f}}}^0
\frac{1}{\alpha_\chi} \, d\ln\frac{\chi}{\chi_\mathrm{f}}
\end{equation}
Now $\alpha_\chi$ is a function of $\psi^2+\chi^2$ and so for $\chi \ll \psi$ is approximately independent of $\chi$, and vice versa.
Therefore, for $t < t_\mathrm{e}$, $\alpha_\chi$ depends on $\ln(\chi/\chi_\mathrm{i})$ and is approximately independent of $\ln\chi_\mathrm{i}$,
and for $t > t_\mathrm{e}$, $\alpha_\chi$ depends on $\ln(\chi/\chi_\mathrm{f})$ and is approximately independent of $\ln\chi_\mathrm{i}$.
Therefore, taking the derivative with respect to $\ln\chi_\mathrm{i}$, we get
\begin{eqnarray}
\frac{d N}{d\ln\chi_\mathrm{i}}
& = & \frac{1}{\alpha_{\chi\mathrm{e}}}
\frac{d\ln\frac{\chi_\mathrm{e}}{\chi_\mathrm{i}}}{d\ln\chi_\mathrm{i}}
- \frac{1}{\alpha_{\chi\mathrm{e}}}
\frac{d\ln\frac{\chi_\mathrm{e}}{\chi_\mathrm{f}}}{d\ln\chi_\mathrm{i}} \\
\label{dndchi}
& = & - \frac{1}{\alpha_{\chi\mathrm{e}}}
\end{eqnarray}
Now
\begin{equation}
\int_{\ln\psi_\mathrm{e}}^{\ln\psi_\mathrm{i}} \frac{d\ln\psi}{\alpha_\psi}
= \int_0^{\ln\frac{\chi_\mathrm{e}}{\chi_\mathrm{i}}}
\frac{1}{\alpha_\chi} \, d\ln\frac{\chi}{\chi_\mathrm{i}}
\end{equation}
and for $t < t_\mathrm{e}$, $\alpha_\psi$ depends on $\ln\psi$ and is approximately independent of $\ln\chi_\mathrm{i}$.
Therefore, taking the derivative with respect to $\ln\chi_\mathrm{i}$, we get
\begin{equation}
- \frac{1}{\alpha_{\psi\mathrm{e}}}
\frac{d\ln\psi_\mathrm{e}}{d\ln\chi_\mathrm{i}}
= \frac{1}{\alpha_{\chi\mathrm{e}}}
\frac{d\ln\frac{\chi_\mathrm{e}}{\chi_\mathrm{i}}}{d\ln\chi_\mathrm{i}}
\end{equation}
and therefore
\begin{equation}
\frac{d\ln\chi_\mathrm{e}}{d\ln\chi_\mathrm{i}}
= \frac{\alpha_{\psi\mathrm{e}}}{\alpha_{\psi\mathrm{e}} + \alpha_{\chi\mathrm{e}}}
\end{equation}
The derivative of Eq.~(\ref{dndchi}) is hence
\begin{eqnarray}
\frac{d^2 N}{(d\ln\chi_\mathrm{i})^2}
& = & - \frac{6\beta_\asymp}{\alpha_{\chi\mathrm{e}}^2(2\alpha_{\chi\mathrm{e}}+3)V_\asymp}
\left[\frac{d\tilde{m}^2_\asymp}{d\beta_\asymp\ln(\psi^2+\chi^2)}\right]_\mathrm{e}
\frac{d\ln\chi_\mathrm{e}}{d\ln\chi_\mathrm{i}} \\
\label{d2ndchi}
& = & - \frac{6\beta_\asymp\alpha_{\psi\mathrm{e}}}{\alpha_{\chi\mathrm{e}}^2
(2\alpha_{\chi\mathrm{e}}+3)(\alpha_{\psi\mathrm{e}}+\alpha_{\chi\mathrm{e}})V_\asymp}
\left[\frac{d\tilde{m}^2_\asymp}{d\beta_\asymp\ln(\psi^2+\chi^2)}\right]_\mathrm{e}
\end{eqnarray}

Now $\chi_\mathrm{i}$ is proportional to the initial angle $\theta$,
therefore, from Eqs.~(\ref{Pext}) and~(\ref{dndchi}),
$P(\ln\theta)$ has a non-trivial extremum at the initial angle $\theta_\mathrm{c}$ that leads to
\begin{equation}\label{a3}
\left.\alpha_{\chi\mathrm{e}}\right|_{\theta_\mathrm{c}} = 3
\end{equation}
or equivalently, from Eq.~(\ref{achi}),
\begin{equation}\label{a32}
\mu^2_\asymp - \left.\tilde{m}^2_{\asymp\mathrm{e}}\right|_{\theta_\mathrm{c}} = 6 V_\asymp
\end{equation}
$\tilde{m}^2_\asymp$ runs from $-\mu_\asymp^2$ at the saddle to $m^2_\asymp$ towards the minimum, so for this to be possible we require
\begin{equation}\label{cons}
\mu^2_\asymp - m^2_\asymp < 6 V_\asymp < 2 \mu^2_\asymp
\end{equation}
or equivalently
\begin{equation}
3 V_\asymp < \mu^2_\asymp < 6 V_\asymp + m^2_\asymp
\end{equation}
$\theta_\mathrm{c}$ can be estimated using Eq.~(\ref{chii}) or Eq.~(\ref{chii2})
\begin{equation}\label{thc}
\theta_\mathrm{c} \sim \exp \left( - \frac{1}{\beta_\asymp}
\int_{\left.\tilde{m}^2_{\asymp\mathrm{e}}\right|_{\theta_\mathrm{c}}}^{m^2_\asymp}
\frac{\alpha_\psi+\alpha_\chi}{2\alpha_\psi}
\left[\frac{d\tilde{m}^2_\asymp}{d\beta_\asymp\ln(\psi^2+\chi^2)}\right]^{-1}
d\tilde{m}^2_\asymp \right)
\end{equation}
and $\sigma$ using Eqs.~(\ref{sigma}), (\ref{d2ndchi}) and~(\ref{a3})
\begin{equation}\label{sigma2}
\sigma = \frac{3}{2\sqrt{\beta_\asymp}\,} \left.
\sqrt{6V_\asymp \left(\frac{\alpha_{\psi\mathrm{e}}+3}{3\alpha_{\psi\mathrm{e}}}\right)
\left[\frac{d\tilde{m}^2_\asymp}{d\beta_\asymp\ln(\psi^2+\chi^2)}\right]_\mathrm{e}^{-1}}
\right|_{\theta_\mathrm{c}}
\end{equation}
with, from Eqs.~(\ref{apsi}) and~(\ref{a32}),
\begin{equation}\label{ape}
\left.\alpha_{\psi\mathrm{e}}\right|_{\theta_\mathrm{c}} =
\frac{3}{2} \, \mathrm{Re} \left( 1 - \sqrt{ 9 - \frac{8\mu_\asymp^2}{3V_\asymp} } \right)
\end{equation}

            \begin{figure}
             \centering

             \mbox{
             {
                 \includegraphics[width=0.47\textwidth]{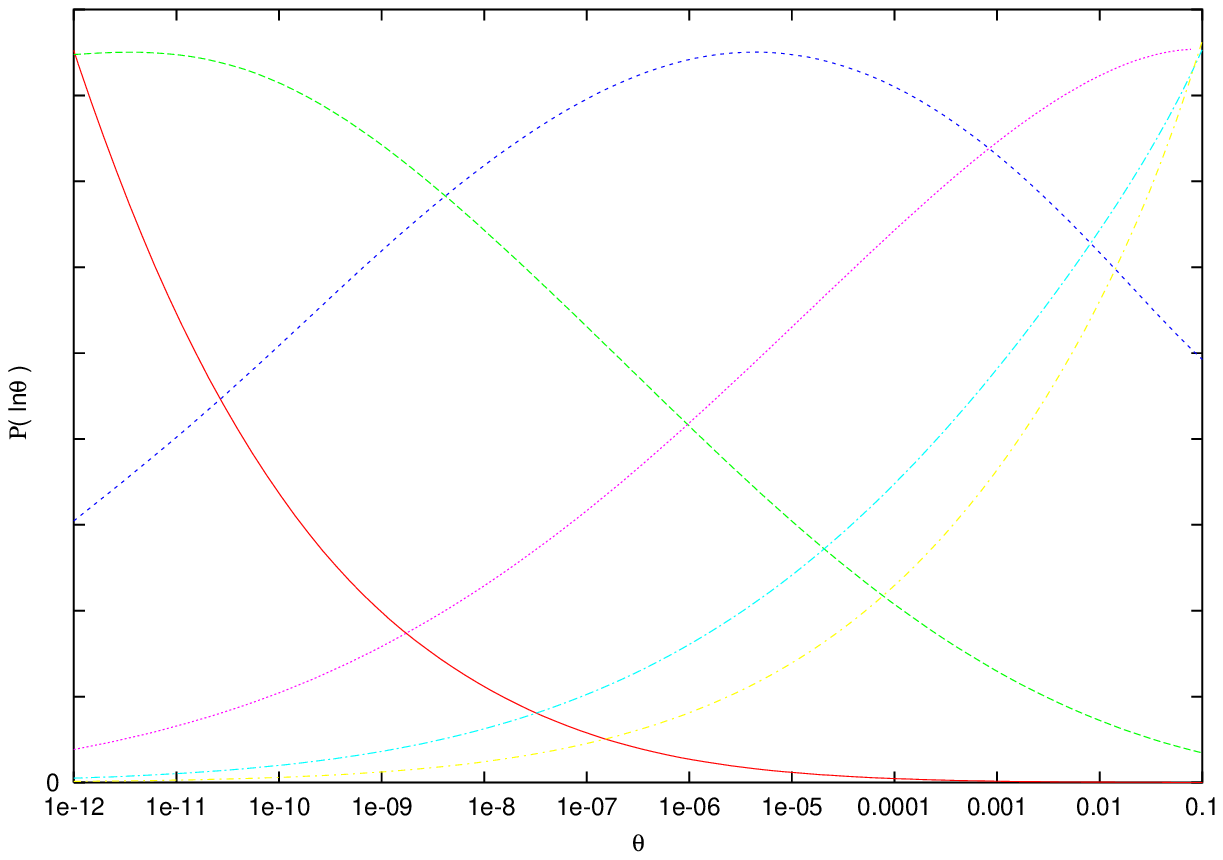}
             }
             \quad
             {
                 \includegraphics[width=0.47\textwidth]{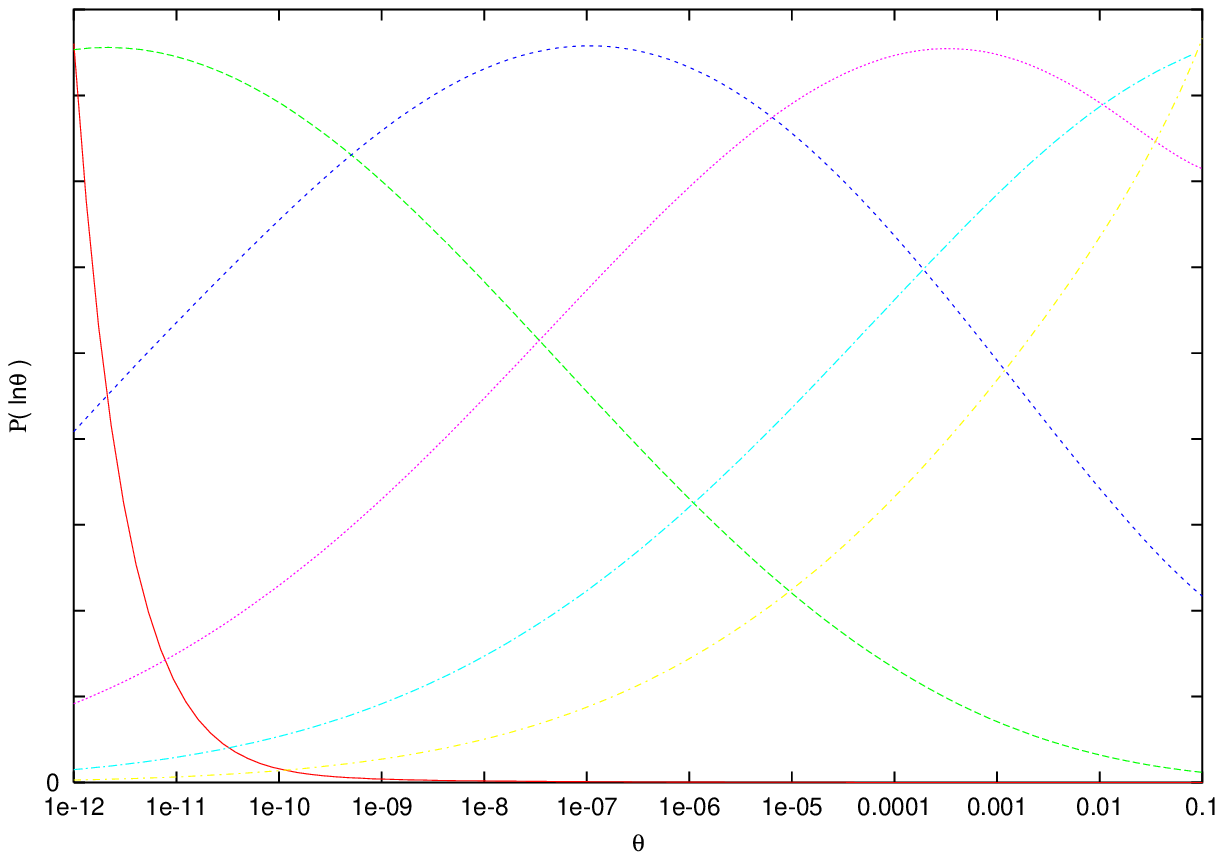}
             }
             
            }

  \mbox{
             {
                 \includegraphics[width=0.47\textwidth]{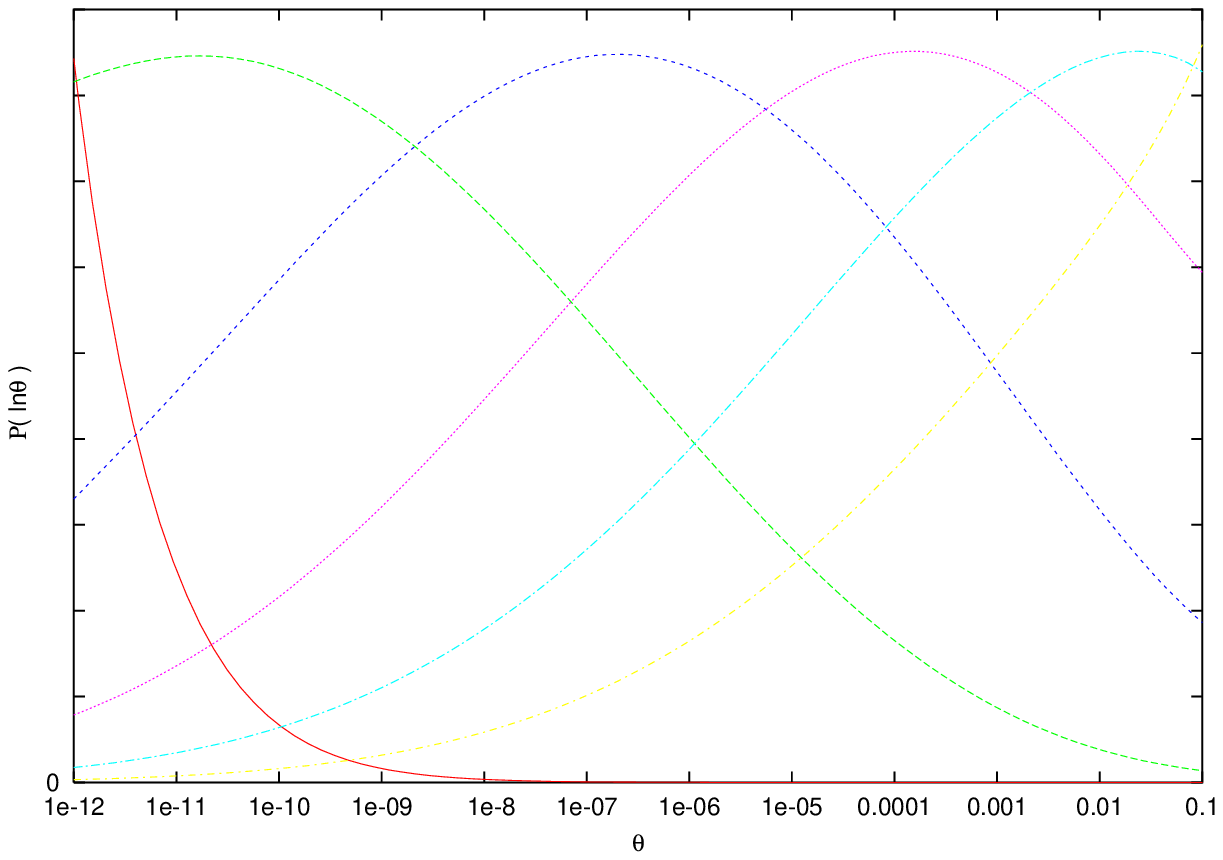}
             }
             \quad
             {
                 \includegraphics[width=0.47\textwidth]{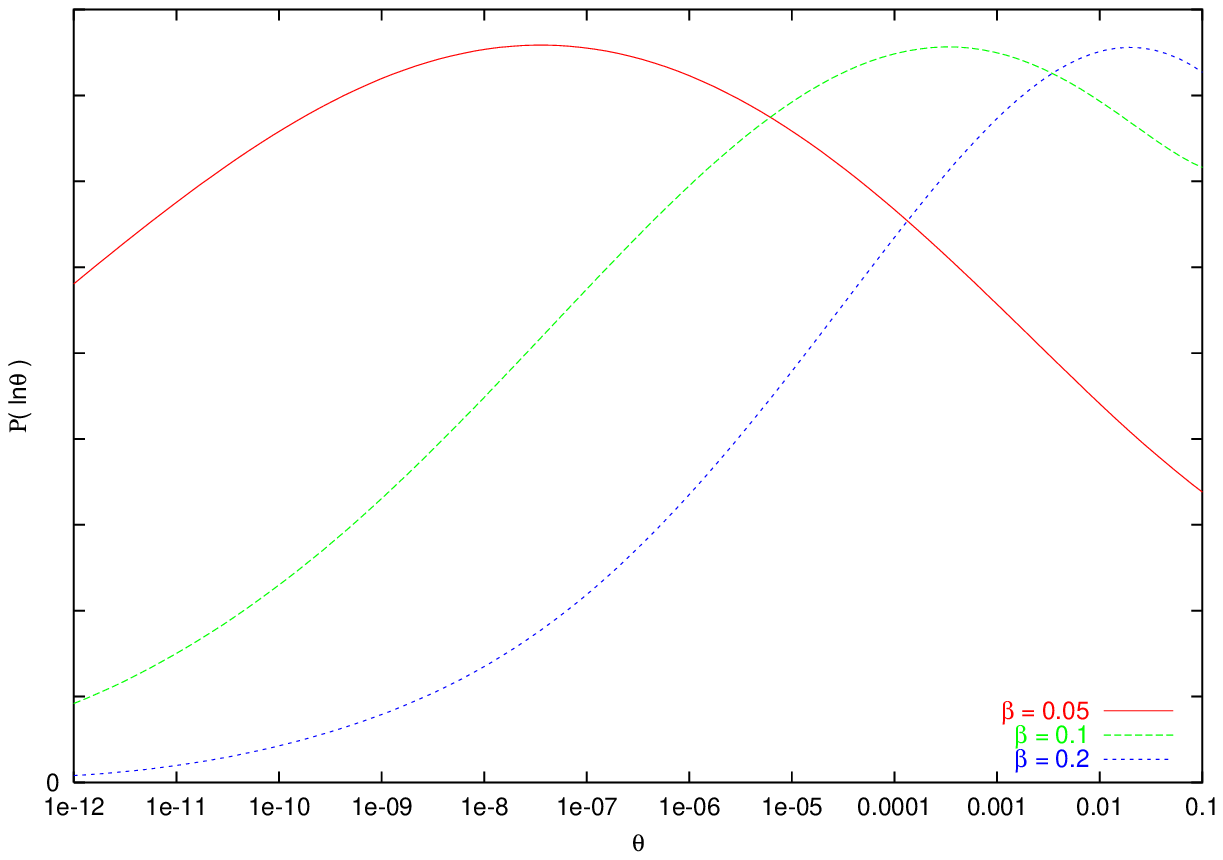}
             }
             
            }

\caption{Log-linear plots of $P(\ln\theta)$ as functions of $\theta$ for $\beta_\asymp = 0.1$,
(top-left) $\nu = 1$, (top-right) $\nu = 2$, (bottom-left) $\nu = 3$, and
$m^2_\smallfrown/V_\smallfrown = 1.5, 2.0, \ldots, 4.0$ from left to right respectively;
and (bottom-right) $\nu = 2$, $m^2_\smallfrown/V_\smallfrown = 3.0$ and
$\beta_\asymp = 0.05, 0.1, 0.2$ from left to right respectively.
$P$ should be normalized so that the area under a graph is one.}
     \label{figP} 
     \end{figure}


For our toy potential, the constraint of Eq.~(\ref{cons}) becomes
\begin{equation}
\begin{array}{ccccccc}
1.69 V_\smallfrown & < & m^2_\smallfrown & < & 3.38 V_\smallfrown & \mbox{for} & \nu = 1 \\
1.49 V_\smallfrown & < & m^2_\smallfrown & < & 3.97 V_\smallfrown & \mbox{for} & \nu = 2 \\
1.29 V_\smallfrown & < & m^2_\smallfrown & < & 4.29 V_\smallfrown & \mbox{for} & \nu = 3
\end{array}
\end{equation}
and, from Eq. (\ref{massrel}), 
\begin{equation}
\frac{d\tilde{m}^2_\asymp}{d\beta_\asymp\ln(\psi^2+\chi^2)}
= \frac{1}{2} m^2_\smallfrown
\end{equation}
Also, to have $0 < \left.\alpha_{\psi\mathrm{e}}\right|_{\theta_\mathrm{c}} < \frac{3}{2}$
requires $3 V_\asymp < \mu_\asymp^2 < \frac{9}{8} (3 V_\asymp)$,
which is a small part of parameter space, and furthermore $\alpha_{\psi\mathrm{e}}$ is the minimum value of $\alpha_\psi$ during the dynamics, so in most cases we can take $\alpha_\psi = \frac{3}{2}$.
It is now straightforward to apply our analytic formulae, and the results agree well with our numerical results shown in Fig.~\ref{figP}.


\subsection{Power spectrum}
\label{ps}

Our final power spectrum is given by Eq.~(\ref{formula})
\begin{equation}
\mathcal{P}_\mathcal{R} =
\left( \frac{\partial N}{\partial \phi} \right)^2 \mathcal{P}_{\delta\phi}
+ \left( \frac{\partial N}{\partial \theta} \right)^2 \mathcal{P}_{\delta\theta}
\end{equation}
with our analytic formulae for the right hand side given by Eqs.~(\ref{note}), (\ref{apxc}), (\ref{dndth}), (\ref{thc}) and (\ref{sigma2}).
For scales which left the horizon when $\ln(\phi/\phi_0) \ll 1$, these results have the form
\begin{equation}\label{Psim}
\mathcal{P}_\mathcal{R} \sim \left(\frac{H_0}{2\pi\phi_0}\right)^2 \left[
A \left(\frac{k}{a_\star H_0}\right)^{-2\alpha}
+ \left( \frac{\partial N}{\partial \theta} \right)^2 \right]
\end{equation}
with $H_0 \sim M_\mathrm{susy}^2 \lesssim 10^{-16}$,
\begin{equation}
\phi_0 \sim \exp \left( - \frac{B}{\beta_\smallfrown} \right)
\end{equation}
and
\begin{equation}\label{Nbeta}
\frac{\partial N}{\partial\theta} \sim
- \exp \left( \frac{C}{\beta_\asymp} \pm \frac{D}{\sqrt{\beta_\asymp}\,} \right)
\end{equation}
where $\beta_\smallfrown \sim \beta_\asymp \sim 10^{-1}$ are one loop renormalization coefficients, $A$, $B$, $C$ and $D$ are constants of order one, and we have chosen the evaluation point $\star$ such that $\ln(\phi_\star/\phi_0) \sim 1$.
The main parameter of our model is $\alpha$, which is defined in Eq.~(\ref{alpha}), and from Eqs.~(\ref{mcon1}) and~(\ref{mcon2}), we expect
\begin{equation}
0.1 \lesssim \alpha < 3
\end{equation}

On large scales the spectrum of Eq.~(\ref{Psim}) is dominated by the power law spectrum $\propto k^{-2\alpha}$ from the radial fluctuations, then, over a wide range of scales
\begin{equation}\label{krange}
\Delta\ln k \sim \frac{1}{\alpha} \ln\left|\frac{\partial N}{\partial \theta}\right|
\end{equation}
it is dominated by the flat spectrum from the angular fluctuations, until finally the spectrum drops off on scales that left the horizon after $\ln(\phi/\phi_0) \sim 1$.
One then expects some tens more $e$-folds of inflation as $\phi$ rolls out via the saddle to its minimum at Planckian values, and possibly ten more due to thermal inflation.

\begin{figure}[h]
\begin{center}    
\epsfxsize = 1.0\textwidth
\epsffile{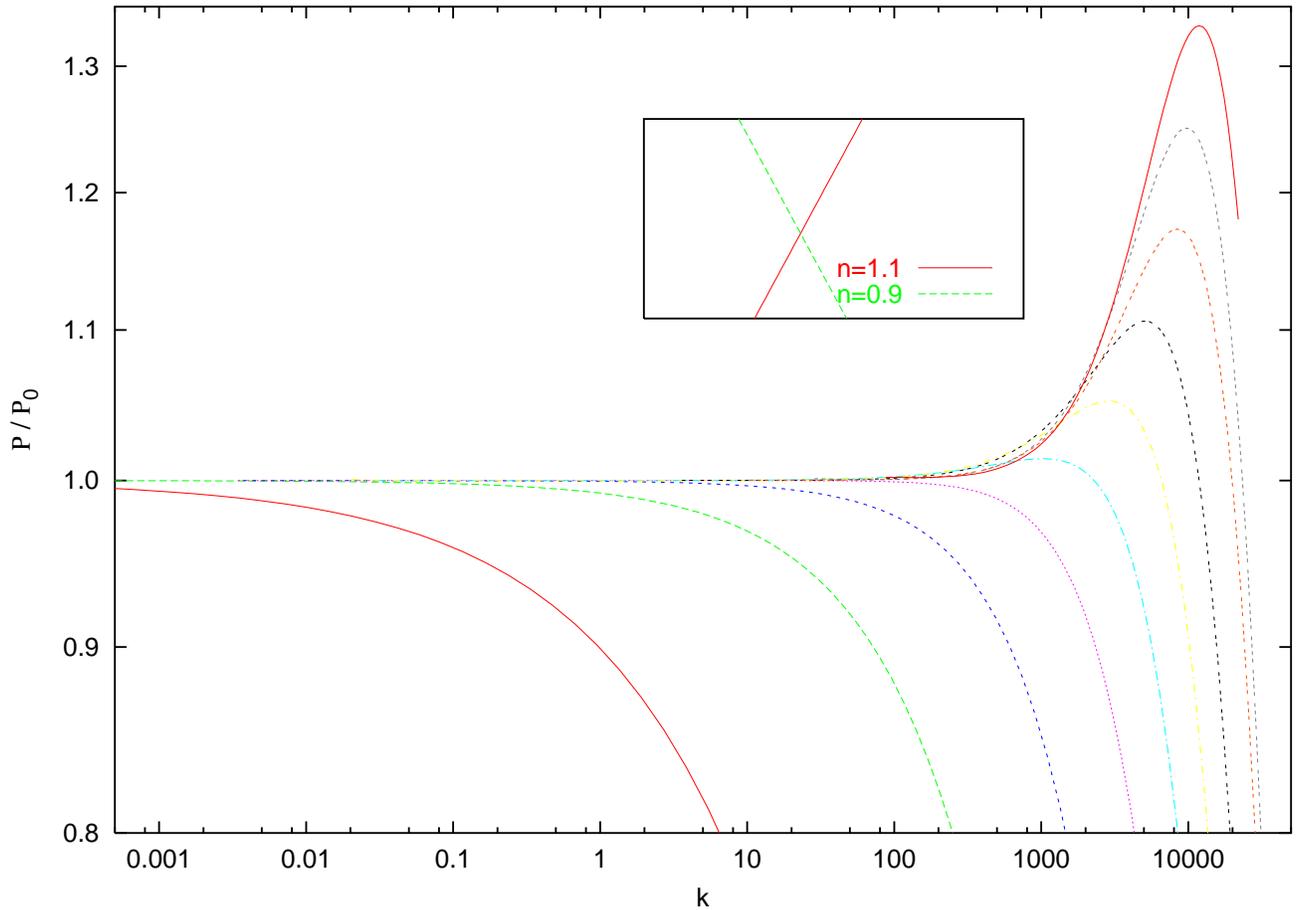}
\end{center}        
\caption{\label{pfig}
Log-log plot of numerical results for $\mathcal{P}/\mathcal{P}_0$ as a function of $k$ for $\alpha = 0.4, 0.6, \ldots,  2.2$, from left to right respectively.
We fix $a$ and hence $k$ by taking $aH = 10^4$ at $\alpha\ln(\phi/\phi_0) = 1$ so that the mode $k=10^4$ left the horizon at $\alpha\ln(\phi/\phi_0) = 1$.
The observable scale corresponding to a given value of $k$ is model dependent.
We show the slopes corresponding to spectral indices $n=1.1$ and $n=0.9$ for comparison.}
\end{figure}

As we saw in Section~\ref{ntheta}, and see from Eq.~(\ref{Nbeta}), we can reasonably expect $\partial N/\partial \theta$ to be exponentially large, so that the range of scales in Eq.~(\ref{krange}) can be more than wide enough to cover observable scales.
The tens of $e$-folds of inflation after $\ln(\phi/\phi_0) \sim 1$, which is a significant fraction of the $40$ odd required for inflation at $H_0 \sim 10^{-16}$, is then likely to put observable scales within the range of scales of Eq.~(\ref{krange}).
Over this range, the spectrum has amplitude
\begin{equation}
\mathcal{P}_\mathcal{R} = \left(\frac{H_0}{2\pi\phi_0}\right)^2 \left( \frac{\partial N}{\partial \theta} \right)^2
\end{equation}
and, from Eq.~(\ref{apxc}), our analytic formula for the spectral index is
\begin{eqnarray}
n_\mathcal{R} - 1 & \simeq & \frac{d\ln\mathcal{P}_{\delta\theta}}{d\ln k}
\\ \label{nr} & = &
- 2 \left[ 2^\alpha \cos\left(\frac{\pi\alpha}{2}\right) \frac{\Gamma(2-\alpha)}{1+\alpha}
+ \frac{8 x_\mathrm{c}^{2-\alpha}}{3 \alpha (2-\alpha)} \right]
\alpha \ln\frac{\phi_\star}{\phi_0} \left(\frac{k}{a_\star H_0}\right)^\alpha
\end{eqnarray}
Numerical results for $\mathcal{P}_{\delta\theta}$ are given in Fig.~\ref{pfig}.
The number of $e$-folds after $\ln(\phi/\phi_0) \sim 1$ determines whether the running of the spectrum shown in Eq.~(\ref{nr}) and Fig.~\ref{pfig} will be on the smallest observable scales or on scales too small to observe, which is heavily model dependent.

\section{Observations}
\label{obs}

\subsection{Observable predictions}

Let us briefly summarize the observable predictions of our model.
We expect a spectrum with negligible deviation from scale invariance over a wide range of $k$ with running becoming significant at (very) small scales.
The asymptotic form of the power spectrum on larger scales is
\begin{equation}
n_\mathcal{R} - 1 = \left\{
\begin{array}{ccc}
- A k^\alpha & \mbox{for} & 0 < \alpha < 1 \\
+ A k^\alpha & \mbox{for} & 1 < \alpha < 2 \\
+ A k^2 & \mbox{for} & \alpha > 2
\end{array}
\right.
\end{equation}
where $A$ is a positive constant and $\alpha$, which is the main parameter which characterizes our model, is expected to be in the range $0.1 \lesssim \alpha < 3$.
The value of $A$ affects the scales at which the deviation from scale invariance becomes significant, and depends on both inflationary parameters and the post-inflationary history and so is highly model dependent.
The full spectra including the characteristic behavior of the deviation from scale invariance at small scales are shown in Fig.~\ref{pfig}.

\subsection{Caution on searching for running}

The usual approach to analyzing the running of the spectrum is based on the standard slow-roll hierarchy $|n''| \ll |n'| \ll |n-1| \ll 1$ and neglects the running of the running $n''$. This is not justified from either theory or observations \cite{gsr,scott}.
In this respect, the general form of the spectral index we obtained in our discussion of cosmic perturbations
\begin{equation}\label{parame}
n-1 = A k^\alpha
\end{equation}
would be better motivated to be used for a test of non-zero running rather than the usual constant running, $n''=0$, approach.
Its running
\begin{equation}
\frac{dn}{d\ln k} = \alpha A k^\alpha
\end{equation}
illustrates our theoretical expectations that the hierarchy $|n''| \ll |n'| \ll |n-1| \ll 1$ is valid only for a limiting region of parameter space, $\alpha \ll 1$, where the running would in any case be small, while for a wider range of $\alpha$ the running is significant but so is the running of the running, $|n''| \sim |n'| \sim |n-1|$.

It is also worth noting that Eq.~(\ref{parame}) indicates that running and running of running are expected to be most significant towards smaller scales, \mbox{i.e.} negligible $n'$ at large scales does not necessarily guarantee the absence of running in the spectrum, which cannot be taken into account if we ignore $n''$.
Thus it is crucial for our observations to probe the smallest possible scales to search for a signal of running.

\section{Discussion and conclusions}
\label{conc}

We investigated inflation on a simple one complex dimensional moduli space and found that the inflationary dynamics is surprisingly rich.
Both real dimensions as well as multiple points of enhanced symmetry play an essential role.
Simplifying to a single real dimensional model or neglecting the effects of the points of enhanced symmetry would lead to completely different expectations for the spectrum of curvature perturbations produced during the inflation.
Interestingly, while such over simplified models tend to produce observationally inviable spectra, we have shown that, for large regions of parameter space, inflation on a simple one complex dimensional moduli space produces an almost exactly flat spectrum for a wide range of scales consistent with current observations.
On smaller, possibly observable, scales running of the spectrum becomes significant providing the opportunity for the model to be tested by future observations.

\subsection*{Acknowledgements}
E.D.S. thanks Misao Sasaki and Takahiro Tanaka for helpful discussions, and K.K. thanks Joanne D. Cohn for helpful advice and continuous encouragement.
We also thank the SF03 Cosmology Summer Workshop for hospitality while this work was in progress.
E.D.S. was supported in part by
the Astrophysical Research Center for the Structure and Evolution of the Cosmos
funded by the Korea Science and Engineering Foundation and the Korean Ministry of Science,
and by the Korea Research Foundation grant KRF PBRG 2002-070-C00022.
K.K. was supported in part by NSF under grant AST-0205935.

\end{document}